\PassOptionsToPackage{unicode}{hyperref}
\PassOptionsToPackage{hyphens}{url}
\documentclass[
]{article}
\usepackage{xcolor}
\usepackage{amsmath,amssymb}
\usepackage{iftex}
\ifPDFTeX
  \usepackage[T1]{fontenc}
  \usepackage[utf8]{inputenc}
  \usepackage{textcomp} % provide euro and other symbols
\else % if luatex or xetex
  \usepackage{unicode-math} % this also loads fontspec
  \defaultfontfeatures{Scale=MatchLowercase}
  \defaultfontfeatures[\rmfamily]{Ligatures=TeX,Scale=1}
\fi
\usepackage{lmodern}
\ifPDFTeX\else
\fi
\IfFileExists{upquote.sty}{\usepackage{upquote}}{}
\IfFileExists{microtype.sty}{% use microtype if available
  \usepackage[]{microtype}
  \UseMicrotypeSet[protrusion]{basicmath} % disable protrusion for tt fonts
}{}
\makeatletter
\@ifundefined{KOMAClassName}{% if non-KOMA class
  \IfFileExists{parskip.sty}{%
    \usepackage{parskip}
  }{% else
    \setlength{\parindent}{0pt}
    \setlength{\parskip}{6pt plus 2pt minus 1pt}}
}{% if KOMA class
  \KOMAoptions{parskip=half}}
\makeatother
\usepackage{longtable,booktabs,array}
\usepackage{calc} % for calculating minipage widths
\usepackage{etoolbox}
\makeatletter
\patchcmd\longtable{\par}{\if@noskipsec\mbox{}\fi\par}{}{}
\makeatother
\IfFileExists{footnotehyper.sty}{\usepackage{footnotehyper}}{\usepackage{footnote}}
\makesavenoteenv{longtable}
\usepackage{graphicx}
\makeatletter
\newsavebox\pandoc@box
\newcommand*\pandocbounded[1]{% scales image to fit in text height/width
  \sbox\pandoc@box{#1}%
  \Gscale@div\@tempa{\textheight}{\dimexpr\ht\pandoc@box+\dp\pandoc@box\relax}%
  \Gscale@div\@tempb{\linewidth}{\wd\pandoc@box}%
  \ifdim\@tempb\p@<\@tempa\p@\let\@tempa\@tempb\fi% select the smaller of both
  \ifdim\@tempa\p@<\p@\scalebox{\@tempa}{\usebox\pandoc@box}%
  \else\usebox{\pandoc@box}%
  \fi%
}
\def\fps@figure{htbp}
\makeatother
\usepackage{bookmark}
\IfFileExists{xurl.sty}{\usepackage{xurl}}{} % add URL line breaks if available
\hypersetup{
  hidelinks,
  pdfcreator={LaTeX via pandoc}}

\title{Scheduling the Unschedulable: Taming Black-Box LLM Inference at Scale}
\author{
  Renzhong Yuan, Yijun Zeng, Xiaosong Gao, Linxi Yu, Haochun Liao, Han Wang \\
  SotaLab, Zhihu Inc.
}
\date{}

\begin{document}

\maketitle

\begin{abstract}
When output token counts can be predicted at submission time (Gan et
al., 2026), client-side scheduling against a black-box LLM API becomes
semi-clairvoyant: decisions condition on coarse token priors even though
the provider's internals remain hidden. We decompose this boundary
problem into three separable concerns: allocation (inter-class share via
adaptive DRR), ordering (intra-class sequencing with feasible-set
scoring), and overload control (explicit admit/defer/reject on a cost
ladder). An information ladder experiment shows that coarse magnitude
priors---not class labels alone---are the practical threshold for useful
client control; removing magnitude inflates short-request P95 by up to
5.8× and degrades deadline satisfaction. Under balanced / high
congestion the full stack achieves 100\% completion, 100\% deadline
satisfaction, and useful goodput of 4.2±1.6 SLO-meeting requests/s with
short P95 within tens of milliseconds of quota-tiered isolation. A
predictor-noise sweep confirms graceful degradation under up to 60\%
multiplicative error. Heavy-dominated regimes separate policies on
completion, tail, and interpretable shedding. We further compare
short-priority allocation (biased toward interactive traffic) with Fair
Queuing (round-robin across classes): Fair Queuing achieves +32\%
short-request P90 improvement over FIFO with only +17\% long-request
overhead, versus Short-Priority's +27\% / +116\%
trade-off---demonstrating that the allocation layer accommodates
different fairness objectives without changing the remaining stack. We
contribute the three-layer client-side decomposition, controlled
evaluation of joint metrics across regimes, allocation-policy
alternatives, and overload-policy evidence linking cost-ladder shedding
to the stated service objective.
\end{abstract}

\section{Introduction}\label{introduction}

LLM inference is increasingly consumed through hosted APIs whose serving
engines are opaque to the client: batching, caching, and internal
scheduling are not observable, and in-flight requests cannot be
reordered or preempted from the client side. The client's leverage is
therefore narrow but real---it can choose when to admit work, how to
partition send opportunities across classes, and in what order to
release requests.

For a long time, that leverage offered little scheduling structure. If
output length is unknown at submission time, then so is the dominant
component of per-request cost; without a defensible notion of ``size,''
classical size-aware or weighted fair intuitions (Parekh and Gallager,
1993) do not transfer to the black-box boundary, and client policies
lack a workload unit on which to allocate or admit.

Gan et al.~(2026) demonstrate that output-length prediction under hybrid
LLM demand is accurate enough that scheduling can treat token counts as
a first-class input. Under that premise, each pending request can be
associated with a coarse prior on cost before it is sent. The
client-side problem is still not clairvoyant---priors are noisy and the
provider's internal state remains hidden---but it is semi-clairvoyant in
the usual scheduling sense: decisions can condition on estimated work.
SageSched addresses demand uncertainty and hybridity inside a
schedulable serving context; here we ask what becomes possible when
similar coarse magnitude information is available only at the client, in
front of an opaque engine. We treat SageSched's finding as the enabling
foundation for this boundary question, not as a contribution of this
paper.

Semi-clairvoyance alone does not make the problem easy. Offered load
affects per-request latency; long jobs can still block short ones unless
arrivals are shaped; and if some requests are dropped, deferred, or
canceled, raw tail latency among completions can improve for the wrong
reason. Following the goodput-oriented perspective on generative serving
(Zhong et al., 2024) and SLO-aware prediction-serving practice
(Crankshaw et al., 2017; Romero et al., 2021), we judge client policies
jointly on short-tail behavior, completion and deadline satisfaction,
and useful goodput---throughput counting only finished, SLO-meeting
work.

Once priors exist, the natural client control plane separates into three
analytically separable concerns: allocation (inter-class share of send
opportunities over time), ordering (intra-class choice of the next
request to release), and overload control (explicit admission versus
pushing all work into the provider and accepting implicit failures).
This three-way decomposition matches how operators already talk about
fairness, head-of-line risk, and admission---and it yields interfaces
that can be tuned and diagnosed independently when joint metrics shift.

\textbf{Contributions.} (1) We give the first systematic formulation of
token-predictable client-side scheduling at the black-box API as a
semi-clairvoyant arrival-shaping problem, building on output-length
predictability as the enabling premise (Gan et al., 2026). (2) We
decompose the policy space into allocation, ordering, and overload
control, tying each layer to a distinct control object and failure mode.
(3) We provide a concrete three-layer instantiation and a controlled
evaluation in a congestion-aware mock provider across four
regimes---balanced / medium, balanced / high, heavy-dominated / medium,
and heavy-dominated / high---with five random seeds per scenario,
mapping joint metrics as a function of regime. The evaluation includes a
four-level information ladder (no-information blind, class-only routing,
coarse semi-clairvoyant priors, oracle upper bound) as foundational
evidence for where coarse magnitude matters under that premise, plus an
overload shedding-policy comparison and a predictor-quality sweep over
multiplicative length error, connecting results to coarse---not
oracle---predictability.

\section{Motivation}\label{motivation}

Before coarse output-size information was available at the client,
black-box scheduling had almost no principled basis. Without a link from
prompt to expected generation length, a client could not rank requests
by predicted load, could not allocate capacity across interactive and
batch classes in proportion to meaningful work units, and could not
reason about which admissions would stress the provider. Policy choices
were therefore dominated by fixed limits and informal rules.

Once requests carry usable size priors, the model changes. The client
can discriminate expensive from cheap work, sequence heavy requests to
reduce predictable head-of-line effects, bias send windows toward
latency-sensitive classes, and decline or defer arrivals when projected
congestion threatens SLOs. The design space is new in the sense that
these knobs become technically meaningful at the API boundary; it is not
a small refinement of prior FIFO practice.

Black-box constraints make that space structurally unusual. The client
does not schedule GPUs; it shapes an arrival process into an opaque
server. Feedback is delayed and aggregate; controls are admission timing
and class-wise release order, not internal preemption. As soon as those
controls are used, tensions appear among (i) protecting short-request
latency, (ii) completing long work, and (iii) producing useful completed
output under deadlines. Improving one dimension by silently starving or
timing out another is always possible; a useful formulation must expose
those couplings.

With token priors in hand, allocation, ordering, and overload become
different control objects: share rules answer ``which class gets the
next send opportunity under congestion?''; sequencing answers ``which
eligible job within a class minimizes predictable head-of-line risk?'';
admission answers ``when should work be deferred or rejected before it
enters the black box?'' Stating them separately keeps failure modes
separable---starvation across classes, blocking within a class, and
saturation that would otherwise surface only as timeouts---and makes the
client plane operationally meaningful under joint metrics. The following
section describes our realization of that structure.

\section{Design}\label{design}

Our scheduler runs entirely on the client. It observes completions,
delays, and queue pressure only as seen through the API; its outputs are
admit/defer/reject decisions and the relative timing of submitted calls.
Coarse token-length priors are assumed for each request before
submission.

\subsection{Three layers}\label{three-layers}

We organize policy into three logical layers:

\begin{enumerate}
\def\labelenumi{\arabic{enumi}.}
\item
  \textbf{Allocation} decides inter-class long-term share: which class
  (e.g., short versus heavy) should receive the next send opportunity
  under work-conserving weighted fair scheduling. We use Deficit Round
  Robin (DRR) with congestion-aware adjustment of class weights so that
  interactive traffic retains protected share when load rises without
  leaving capacity idle when the heavy class is empty. Concretely, each
  class maintains a deficit counter; when selected, a class may send if
  \texttt{deficit\ ≥\ estimated\_cost}, otherwise the deficit
  accumulates. A work-conserving borrowing rule allows an idle class's
  unused quota to be consumed by a backlogged peer. Congestion feedback
  scales weights: under stress the short class's effective share grows,
  biasing send opportunities toward latency-sensitive work.
\item
  \textbf{Ordering} decides intra-class sequencing for the heavy class:
  among requests eligible under fairness constraints, which job to
  submit next. A slowdown-aware, feasible-set rule scores candidates
  using a weighted combination:
  \texttt{score\ =\ w₁·(wait\ /\ cost)\ −\ w₂·(size\ /\ ref)\ +\ w₃·urgency},
  where \texttt{wait} is queue residence time, \texttt{cost} is the
  token prior, and \texttt{urgency} captures deadline proximity. The
  formula favors older and smaller jobs while respecting urgency,
  reducing predictable head-of-line blocking. Across all runs we report,
  we observed zero violations of the ordering layer's feasibility
  constraints.
\item
  \textbf{Overload control} sits at the admission boundary. It
  integrates observable signals---recent completion latency, outstanding
  in-flight work, and tail behavior---into a severity score:
  \texttt{severity\ =\ w\_load·provider\_load\ +\ w\_queue·queue\_pressure\ +\ w\_tail·tail\_latency\_ratio}.
  Above progressive thresholds (\texttt{t₁\ =\ 0.45} for defer,
  \texttt{t₂\ =\ 0.65} for reject-xlong, \texttt{t₃\ =\ 0.80} for
  reject-long) it maps estimated cost to defer/reject decisions through
  a cost ladder with bucket weights: medium = 0, long = 1, xlong = 2.
  Short requests are never rejected. The intent is to replace implicit
  timeout failures with explicit, objective-aligned shedding: sacrifice
  is visible in the client policy and concentrated on expensive work,
  matching interactive service semantics under the joint metrics in §4.
  §4.6 contrasts this ladder with class-agnostic and stress-reversed
  alternatives.
\end{enumerate}

The allocation layer selects a class; the ordering layer names a
concrete request in that class; the overload layer may block or delay
that release. Each layer targets a different pathology: starvation
across classes, blocking within a class, and uncontrolled saturation.

\subsection{Why separate layers}\label{why-separate-layers}

Keeping allocation, ordering, and overload control in separate layers is
a design for control clarity and measurable responsibility. Each layer
exposes a smaller policy surface: DRR weights and congestion reaction
for share, a feasible-set score for in-class order, and thresholds plus
a bucket severity map for admission. When metrics move, this
decomposition indicates which decision class to adjust. It also aligns
with separable failure modes: bad share starves a class globally; bad
ordering creates head-of-line blocking within a class; missing overload
control pushes saturation into the provider where it becomes
unobservable client-side. The evaluation therefore reports not only
end-to-end numbers but layerwise and overload-policy contrasts that
stress each concern in isolation where possible.

\subsection{Assumptions}\label{assumptions}

The design assumes (i) priors are coarse but correlated with actual
cost, and (ii) applications can tolerate deferral or rejection signals
from the client. §4.4 establishes empirically---through a four-level
information ladder from no client-side magnitude signal to coarse
semi-clairvoyance and an oracle upper bound---that this formulation
rests on output-length predictability; §4.9 then sweeps graded
multiplicative error on coarse priors while holding mock physics fixed.
Remaining external-validity limits are discussed in §5. Figure 1
summarizes data flow through the layers.

\begin{figure}
\centering
\pandocbounded{\includegraphics[keepaspectratio,alt={Data path for the client-side stack: allocation, ordering, and overload control ahead of the mock black-box API; coarse length priors are available before dispatch.}]{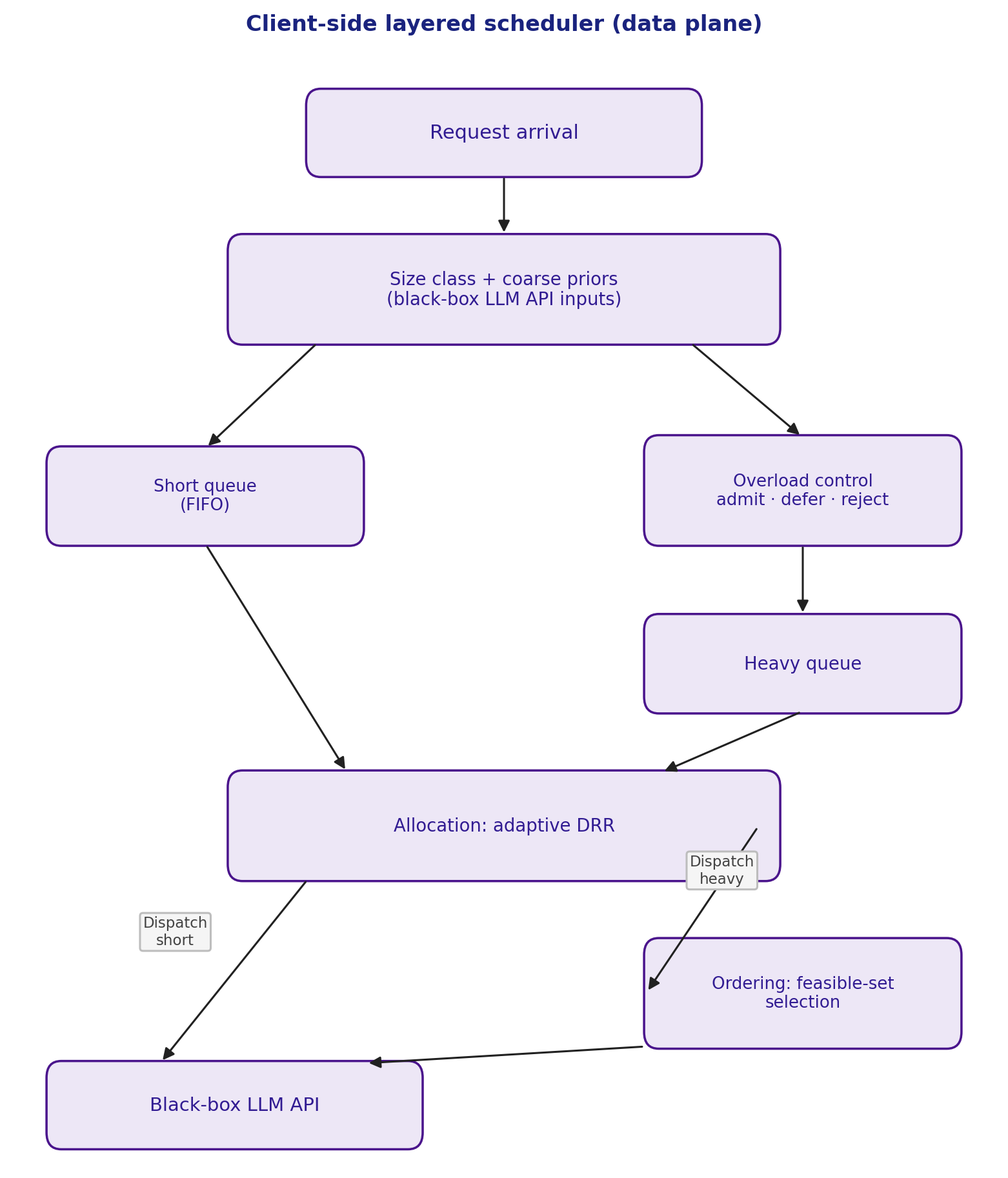}}
\caption{Data path for the client-side stack: allocation, ordering, and
overload control ahead of the mock black-box API; coarse length priors
are available before dispatch.}
\end{figure}

\section{Evaluation}\label{evaluation}

We proceed in four passes: §§4.1--4.3 define the mock provider, regimes,
and joint metrics; §4.4 tests the enabling premise with a four-level
information ladder (fixed Final OLC); §§4.5--4.8 compare structured
policies under coarse priors---including allocation-policy alternatives
(§4.6 Fair Queuing vs Short-Priority) and overload shedding shapes
(§4.7)---and show layerwise progression; §§4.9--4.10 perturb controller
thresholds and predictor noise. §4.4 and §4.10 together bound what
``coarse, not oracle'' requires in this study.

\subsection{Congestion-aware mock provider}\label{congestion-aware-mock-provider}

Real hosted APIs couple client decisions with unobservable server
policies and time-varying contention. To preserve a clear causal
chain---arrival shaping → offered load → load-dependent slowdown →
completions---we use a congestion-aware mock provider whose service
times scale with predicted token work and whose per-request delay grows
with concurrent load. Following simulation methodology established in
recent LLM serving research (Zhong et al., 2024; Agrawal et al., 2024),
the mock is an abstraction, not a faithful model of a specific vendor;
it is chosen so that experiments isolate client-side scheduling while
retaining the qualitative physics this paper cares about: bigger jobs
cost more, overload hurts everyone, and shaping arrivals changes tails
and throughput.

\textbf{Latency calibration.} To ground the mock's per-token cost
assumption, we measured single-request latency versus output tokens on a
production LLM API under low-load conditions (Volcengine Doubao, 18
requests spanning three token buckets). A linear fit yields latency\_ms
= 3294 + 18.7 × output\_tokens with R² = 0.97, confirming that
generation time scales linearly with output length---the key property
the mock relies on. Table 3 summarizes bucket-wise statistics (CSV:
\texttt{paper\_results/tables/latency\_calibration.csv}).

\begin{table}[h]
\centering
\small
\begin{tabular}{lrrrrr}
\toprule
Bucket & Count & Mean tokens & Std tokens & Mean latency (ms) & Std latency (ms) \\
\midrule
medium & 3 & 155 & 35 & 4916 & 608 \\
long & 5 & 670 & 259 & 14968 & 3950 \\
xlong & 10 & 2839 & 907 & 57251 & 16064 \\
\bottomrule
\end{tabular}
\caption{Latency calibration by bucket (Table 3).}
\end{table}

\textbf{Real-trace validation.} To verify that synthetic workload
distributions do not overfit our conclusions, we replay a
ShareGPT-derived output-token distribution against the same mock
provider. ShareGPT-English (388,246 assistant responses from 50k
conversations) exhibits a real-world bucket split: 12\% short (≤64
tokens), 42\% medium (65--256), 46\% long (257--1024), and \textless1\%
xlong (\textgreater1024)---substantially different from our balanced
(50/25/15/10) and heavy-dominated (20/20/30/30) synthetic mixes. Table 6
summarizes key results under high congestion.

\begin{table}[h]
\centering
\small
\begin{tabular}{lllll}
\toprule
Strategy & Short P95 (ms) & Global P95 (ms) & Makespan (ms) & Satisfaction \\
\midrule
direct\_naive & 1449±392 & 25985±1218 & 36352±958 & 0.78±0.04 \\
quota\_tiered & 324±31 & 33337±637 & 41782±623 & 0.77±0.05 \\
\textbf{final\_adrr\_olc} & 330±38 & \textbf{17470±1342} & 22001±54 & \textbf{0.90±0.04} \\
\bottomrule
\end{tabular}
\caption{ShareGPT real-trace validation results under high congestion.}
\end{table}

Under the ShareGPT distribution, final\_adrr\_olc achieves 4.4×
short-P95 improvement over naive dispatch (330 vs 1449 ms), 1.9×
global-P95 reduction over quota-tiered (17470 vs 33337 ms), and +15\%
deadline satisfaction (0.90 vs 0.78 for naive). The qualitative ordering
of policies---and the advantage of structured scheduling under
congestion---holds across both synthetic and trace-derived
distributions, supporting external validity beyond the four synthetic
regimes.

\textbf{Workloads and metrics (preview).} The next two subsections fix
the regimes we stress and the joint metrics we report; §4.4 then asks
whether semi-clairvoyant control is justified at all before we compare
named policies.

\subsection{Workloads and regimes}\label{workloads-and-regimes}

We study two workload mixes (balanced versus heavy-dominated) crossed
with two congestion levels (medium versus high), yielding four regimes:
balanced / medium, balanced / high, heavy-dominated / medium, and
heavy-dominated / high. Requests fall into token buckets (short through
xlong); the client sees a noisy prior for each. Each regime is run with
five independent seeds.

\subsection{Metrics}\label{metrics}

The metrics below are chosen so that tail improvements cannot be read in
isolation from completion and SLO satisfaction.

We report mean ± standard deviation across seeds. Short P95 is the 95th
percentile latency of short requests among completed calls. Global P95
is the corresponding tail over all completed requests. Completion rate
(CR) and deadline satisfaction measure whether work finishes and whether
it finishes within application deadlines. Useful goodput is the rate of
requests that both complete and meet their deadlines---our primary
throughput metric when ``throughput'' must mean useful completed work,
in line with goodput-oriented generative serving (Zhong et al., 2024)
and SLO-grounded serving metrics in prediction systems (Crankshaw et
al., 2017; Romero et al., 2021).

These metrics must be read together. Low global P95 paired with low
completion indicates sacrificed work, not a strictly better system.
Before comparing named policies, §4.4 establishes whether the
semi-clairvoyant formulation itself is empirically justified: the same
three-layer stack is meaningless if the client lacks usable
output-length magnitude.

\subsection{Information ladder: from blind client control to semi-clairvoyant scheduling}\label{information-ladder-from-blind-client-control-to-semi-clairvoyant-scheduling}

Section §3 expresses allocation, ordering, and overload in estimated
token work (DRR weights, slowdown-aware scores with a size term, and
cost-aware admission). Those mechanisms presuppose usable magnitude
priors; without them, the same code reduces to blind budgeting. This
subsection is the evaluation's premise test: we hold the Final (OLC)
stack fixed and vary only what the client may know---no information,
class-only (labels for routing and tiered overload, but neutral
p50/p90), coarse per-request priors, or an oracle upper bound---under
identical mock physics (each job's realized service scale still follows
the generator's nominal workload field \texttt{sim\_workload\_p50}).

We use four information levels on every request:

\begin{enumerate}
\def\labelenumi{\arabic{enumi}.}
\item
  \textbf{No-information blind.} The client has neither per-request
  output-length estimates nor workload-class routing derived from size:
  every request is treated under a single neutral routing bucket with
  fixed neutral p50/p90 for budgeting and scoring. Overload control
  cannot use a long/xlong length ladder; it instead applies a uniform
  admission severity so that deferral and rejection can still track
  aggregate stress without inferring cost from class labels. This
  condition approximates black-box client control when the API exposes
  no reliable length signal.
\item
  \textbf{Class-only.} The generator's class label (short through xlong)
  drives interactive versus heavy routing and tiered overload
  (defer/reject can depend on bucket), but every request still carries
  the same neutral p50/p90 as in no-information blind for allocation
  weights, ordering scores, and token budgets. Intuitively, the client
  knows \emph{which lane} a request uses, not \emph{how large} it is
  within that lane.
\item
  \textbf{Coarse (semi-clairvoyant).} The default setting elsewhere in
  this section: experiment bounds map to per-request p50/p90 used
  throughout allocation, ordering, and budgeting.
\item
  \textbf{Oracle prior (upper bound).} The policy sees the exact mock
  output-token count before dispatch---an information frontier, not a
  deployable predictor.
\end{enumerate}

Across four regimes with five seeds per (regime, condition) cell, each
condition aggregates 20 runs (80 runs total). Figure 2 and Table 1
report mean ± std.

\begin{figure}
\centering
\pandocbounded{\includegraphics[keepaspectratio,alt={Information ladder with Final (OLC) fixed (five seeds per regime × condition). Top: short-request P95 (mean ± std); no-information blind uses red hatching. Bottom: completion rate and useful goodput. Conditions: no-information blind, class-only, coarse semi-clairvoyant, oracle.}]{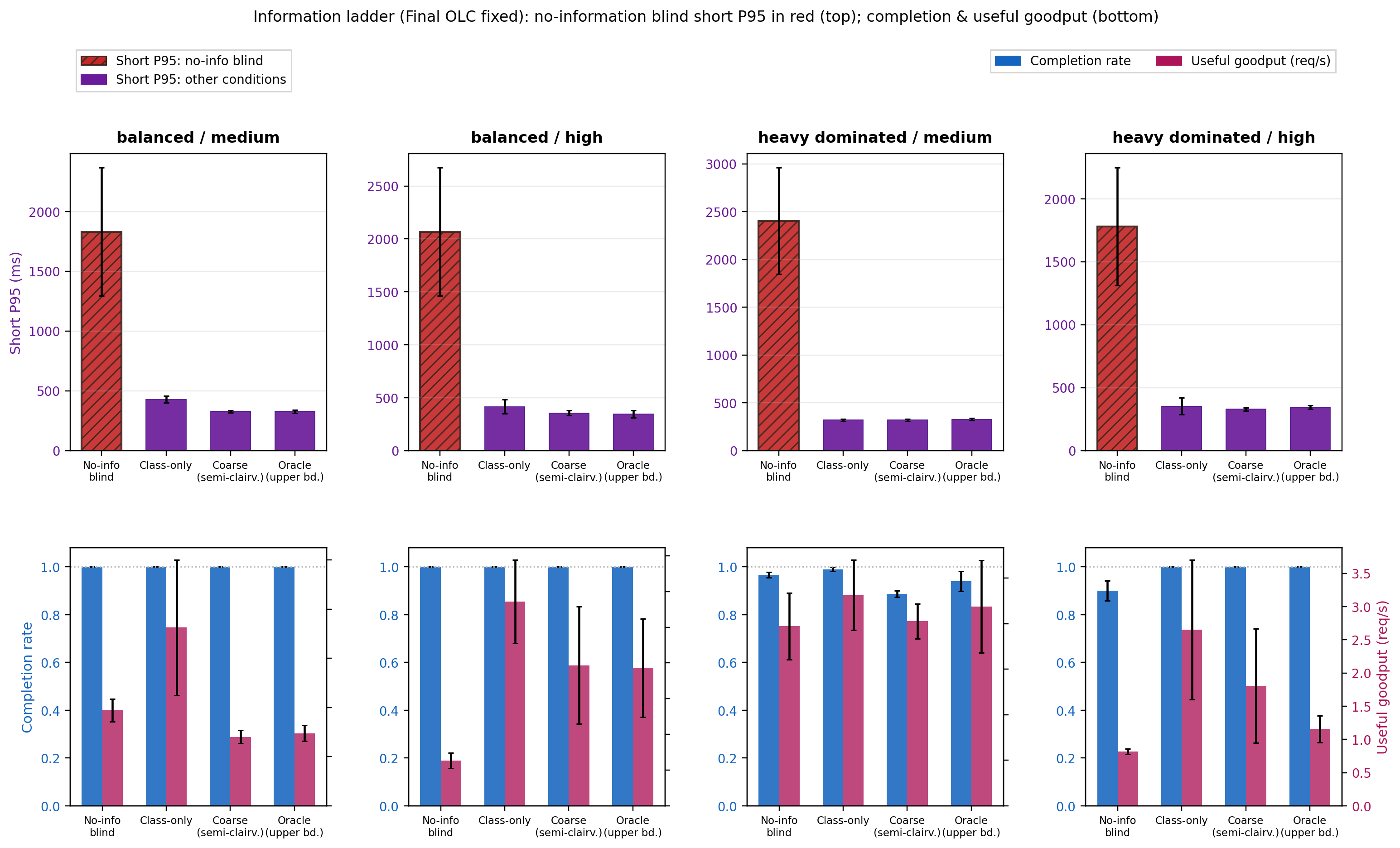}}
\caption{Information ladder with Final (OLC) fixed (five seeds per
regime × condition). Top: short-request P95 (mean ± std); no-information
blind uses red hatching. Bottom: completion rate and useful goodput.
Conditions: no-information blind, class-only, coarse semi-clairvoyant,
oracle.}
\end{figure}

Table 1 summarizes the same runs (CSV:
\texttt{paper\_results/tables/prior\_ablation\_summary.csv}).

\begin{table}[h]
\centering
\footnotesize
\begin{tabular}{llrrrrr}
\toprule
Regime & Information & Short P95 & Global P95 & CR & Satisfaction & Goodput \\
\midrule
\textbf{balanced/medium} & No-info & 1829±536 & 18872±1546 & 1.00 & 0.95±0.10 & 1.9±0.2 \\
& Class-only & 428±29 & 9543±3216 & 1.00 & 1.00 & 3.6±1.4 \\
& Coarse & 327±9 & 21903±2637 & 1.00 & 0.997±0.008 & 1.4±0.1 \\
& Oracle & 327±14 & 19614±2873 & 1.00 & 1.00 & 1.5±0.2 \\
\midrule
\textbf{balanced/high} & No-info & 2067±606 & 26369±3406 & 1.00 & 0.93±0.09 & 1.3±0.2 \\
& Class-only & 415±65 & 7175±989 & 1.00 & 1.00 & 5.7±1.2 \\
& Coarse & 355±23 & 7585±1187 & 1.00 & 1.00 & 3.9±1.6 \\
& Oracle & 346±33 & 7449±1057 & 1.00 & 1.00 & 3.9±1.4 \\
\midrule
\textbf{heavy/medium} & No-info & 2402±556 & 49880±1590 & 0.97±0.01 & 0.77±0.12 & 0.8±0.1 \\
& Class-only & 318±13 & 44610±2501 & 0.99±0.01 & 0.85±0.13 & 0.9±0.2 \\
& Coarse & 318±11 & 47383±1624 & 0.89±0.01 & 0.82±0.07 & 0.8±0.1 \\
& Oracle & 327±13 & 43595±2201 & 0.94±0.04 & 0.83±0.18 & 0.9±0.2 \\
\midrule
\textbf{heavy/high} & No-info & 1779±468 & 46429±5093 & 0.90±0.04 & 0.79±0.06 & 0.8±0.04 \\
& Class-only & 352±67 & 16269±5407 & 1.00 & 0.94±0.08 & 2.7±1.1 \\
& Coarse & 328±13 & 22393±8152 & 1.00 & 0.93±0.06 & 1.8±0.9 \\
& Oracle & 344±15 & 29355±7315 & 1.00 & 0.93±0.08 & 1.2±0.2 \\
\bottomrule
\end{tabular}
\caption{Information ladder results (Table 1).}
\end{table}

\textbf{No-information blind.} Short-request P95 is dramatically higher
than under any informed condition---for example balanced / high 2067±606
ms versus 355±23 ms with coarse priors (roughly 5.8× in mean) and
balanced / medium 1829±536 ms versus 327±9 ms. Deadline satisfaction
falls to 0.93±0.09 and 0.95±0.10 in those balanced cells even when
completion stays 1.00; heavy-dominated / high combines 0.90±0.04
completion, 0.79±0.06 satisfaction, and 1779±468 ms short P95. Useful
goodput stays in the \textasciitilde0.8--1.9 req/s range alongside weak
satisfaction, so the blind condition is not ``healthy'' under the joint
view. In Figure 2 (top row), the red, hatched first bar marks this
condition in every regime.

\textbf{Class-only versus coarse.} With class labels alone, the client
regains routing and bucket-aware overload, but p50/p90 stay neutral, so
allocation, ordering, and budgets cannot distinguish cheap from
expensive work within a class. That is routing structure without
per-request magnitude---and it is not the same as the coarse
semi-clairvoyant target in §3, where share, sequencing, and admission
are token-informed.

\textbf{Why class-only sometimes shows higher goodput.} This apparent
paradox reflects different operating points on the Pareto frontier, not
a failure of coarse priors. Under class-only, neutral budgets
underestimate heavy-request costs, causing the scheduler to admit more
work than it would with accurate magnitude information. The result is
higher throughput of completed requests---but at a cost: short P95
degrades (e.g., 415±65 ms versus 355±23 ms at balanced / high) and the
scheduler operates closer to saturation. Coarse priors, by contrast,
enable the scheduler to anticipate congestion and defer expensive work
before it impacts latency-sensitive traffic. The two conditions occupy
different points on a joint surface trading off throughput against tail
protection. Which point is ``better'' depends on whether the operator
prioritizes raw goodput or short-request latency---both are valid, but
they are not the same objective. In balanced regimes the contrast is
clear: class-only achieves \textasciitilde5.7 versus \textasciitilde3.9
req/s at balanced / high, but coarse priors pull short P95 into the
\textasciitilde328--355 ms band and hold deadline satisfaction at
\textasciitilde1.00. In heavy-dominated / medium, short P95 for
class-only and coarse is similar (\textasciitilde318 ms), yet completion
and satisfaction still differ (0.99±0.01 and 0.85±0.13 versus 0.89±0.01
and 0.82±0.07), which shows the ladder shifting joint outcomes, not a
single latency figure.

\textbf{Oracle.} Oracle priors track coarse on short tails in several
cells and leave the large no-information gap intact; the practical bar
is coarse magnitude, not exact tokens.

Having fixed what the client knows, §4.5 compares how different
structured policies behave under the coarse semi-clairvoyant setting
used for the rest of the evaluation.

\subsection{Main policy comparisons}\label{main-policy-comparisons}

§4.4 fixed the information frontier for the same Final (OLC)
implementation used below. Quota-tiered isolation, adaptive DRR without
overload control, and the full design (adaptive DRR + overload control)
are evaluated under coarse semi-clairvoyant priors as defined
there---the setting in which allocation, ordering, and overload
admission are expressed in meaningful work units. An uncontrolled direct
naive dispatcher appears in the scatter plots for orientation only.
Table 2 summarizes the three structured policies (mean ± std over
seeds). Figure 3 plots short P95 against completion rate (including
direct naive). Figure 4 plots useful goodput against global P95.

\begin{table}[h]
\centering
\footnotesize
\begin{tabular}{llrrrrrr}
\toprule
Regime & Strategy & Short P95 & Global P95 & Makespan & CR & Satisf. & Goodput \\
\midrule
\textbf{bal./med.} & Quota-tiered & 323.5±14.4 & 8524±2415 & 14501±2667 & 0.90 & 0.89±0.01 & 3.8±0.7 \\
& Adaptive DRR & 327.8±13.3 & 22255±2123 & 43329±3759 & 1.00 & 1.00±0.01 & 1.4±0.1 \\
& \textbf{Final (OLC)} & 320.9±4.3 & \textbf{21239±3087} & 42310±3979 & \textbf{1.00} & 1.00±0.01 & \textbf{1.4±0.1} \\
\midrule
\textbf{bal./high} & Quota-tiered & 320.8±19.7 & 8600±1935 & 14484±3131 & 0.97±0.02 & 0.96±0.02 & 4.2±0.9 \\
& Adaptive DRR & 349.3±36.2 & 7495±1416 & 18204±7944 & 1.00 & 1.00 & 3.8±1.7 \\
& \textbf{Final (OLC)} & 347.4±27.5 & \textbf{7786±1328} & 17152±9402 & \textbf{1.00} & 1.00 & \textbf{4.2±1.6} \\
\midrule
\textbf{heavy/med.} & Quota-tiered & 319.8±15.8 & 20911±1174 & 26410±1212 & 0.70 & 0.60±0.02 & 1.4±0.1 \\
& Adaptive DRR & 327.3±14.1 & 44494±2518 & 58896±2234 & 0.88±0.02 & 0.75±0.18 & 0.8±0.2 \\
& \textbf{Final (OLC)} & 320.8±15.2 & \textbf{45199±1912} & 58812±1282 & \textbf{0.92±0.04} & 0.88±0.07 & \textbf{0.9±0.1} \\
\midrule
\textbf{heavy/high} & Quota-tiered & 319.6±12.7 & 20105±1368 & 26383±1271 & 0.89±0.04 & 0.78±0.03 & 1.8±0.1 \\
& Adaptive DRR & 327.6±19.6 & 18556±8483 & 32079±16917 & 1.00±0.01 & 0.96±0.04 & 2.3±1.2 \\
& \textbf{Final (OLC)} & 325.3±23.1 & \textbf{25119±13315} & 34938±15995 & \textbf{0.99±0.01} & 0.92±0.07 & \textbf{2.0±1.2} \\
\bottomrule
\end{tabular}
\caption{Main policy comparison results (Table 2).}
\end{table}

\textbf{Balanced / medium.} Quota-tiered achieves the lowest global P95
and makespan but completes only 90\% of requests with 89±1\% deadline
satisfaction, whereas adaptive DRR and the full design reach 100\%
completion with 100±1\% satisfaction---illustrating that attractive
tails can accompany withheld work. Useful goodput is highest for
quota-tiered (3.8±0.7) in this regime; the full design matches adaptive
DRR on goodput (1.4±0.1) while preserving completion.

\textbf{Balanced / high.} Quota-tiered completes 97±2\% of requests with
96±2\% deadline satisfaction (Table 2); adaptive DRR and the full design
reach 100\% on both. Short P95 for the full design (347.4±27.5 ms) sits
within tens of milliseconds of quota-tiered (320.8±19.7 ms). Useful
goodput is 4.2±1.6 for the full design versus 4.2±0.9 for quota-tiered
and 3.8±1.7 for adaptive DRR alone, so the full stack matches quota on
mean goodput here while closing the completion gap. Compared to adaptive
DRR alone, adding overload control raises useful goodput from 3.8 to 4.2
at 100\% completion, with about 4.6 rejects and 8.8 defers versus none
without overload control---an 8.6\% relative goodput gain in that paired
comparison.

\textbf{Heavy-dominated / medium.} Quota-tiered reports 70\% completion
and 60±2\% satisfaction versus 92±4\% / 88±7\% for the full design;
global P95 and makespan are lower for quota-tiered partly because fewer
heavy jobs finish. Useful goodput is 1.4±0.1 (quota) versus 0.9±0.1
(full design)---a regime where latency-first shedding and
completion-first policies diverge sharply.

\textbf{Heavy-dominated / high.} Completion is near 100\% for adaptive
DRR (1.00±0.01) and 99±1\% for the full design, versus 89±4\% for
quota-tiered. Useful goodput is highest for adaptive DRR (2.3±1.2), with
the full design at 2.0±1.2 and quota-tiered at 1.8±0.1. Global P95 and
makespan split accordingly (18556±8484 ms and 32079±16917 ms for
adaptive DRR versus 25119±13315 ms and 34938±15995 ms for the full
design). The three-layer stack and quota-tiered isolation therefore
occupy different points on a regime-dependent joint surface; §4.6 shows
how overload shedding shape moves that surface when the full design is
held fixed.

\begin{figure}
\centering
\pandocbounded{\includegraphics[keepaspectratio,alt={Short-request P95 versus completion rate (four regimes; mean ± std over seeds). Structured policies at high completion with moderate short tails; naive dispatch skews toward worse short P95 and lower completion under stress.}]{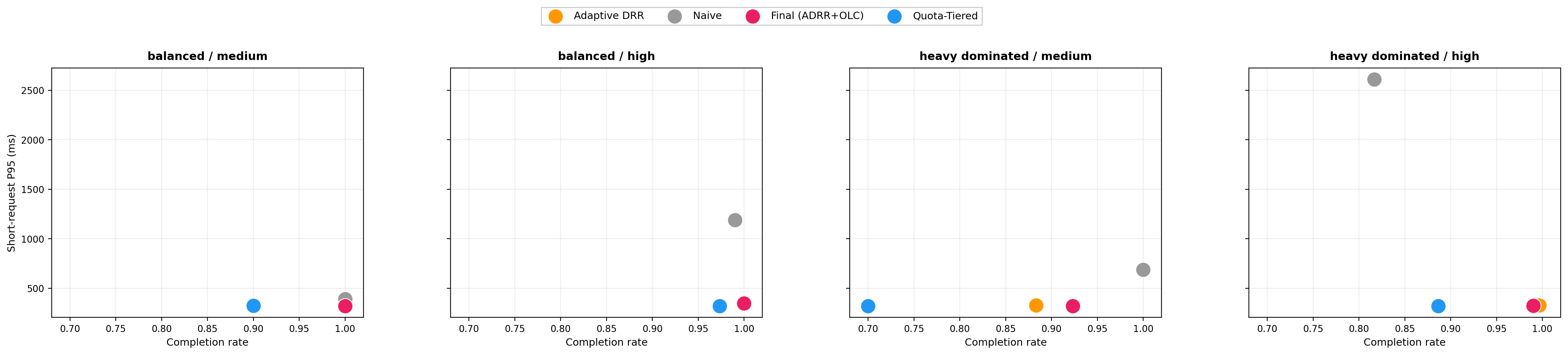}}
\caption{Short-request P95 versus completion rate (four regimes; mean ±
std over seeds). Structured policies at high completion with moderate
short tails; naive dispatch skews toward worse short P95 and lower
completion under stress.}
\end{figure}

\begin{figure}
\centering
\pandocbounded{\includegraphics[keepaspectratio,alt={Useful goodput versus global P95 for the same main-benchmark runs as the previous figure.}]{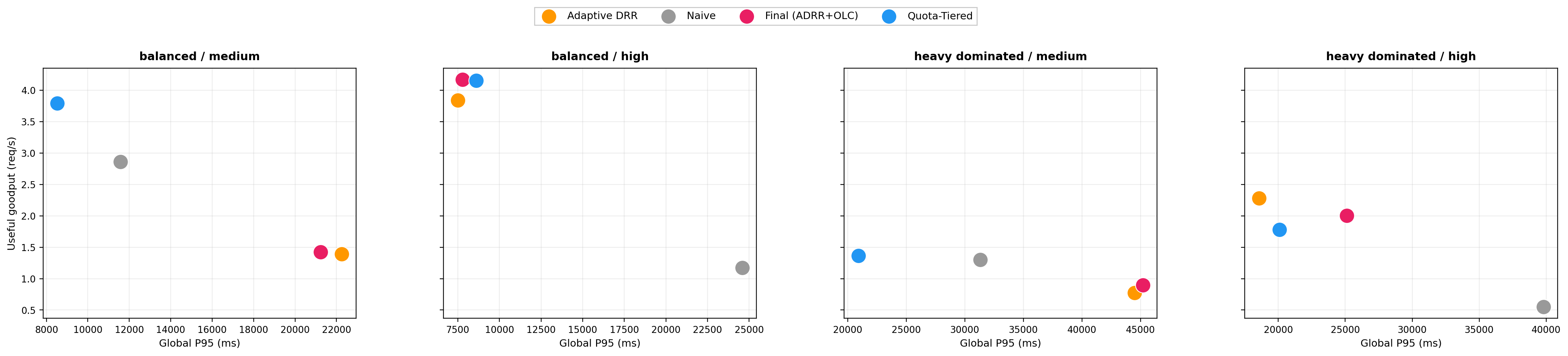}}
\caption{Useful goodput versus global P95 for the same main-benchmark
runs as the previous figure.}
\end{figure}

\subsection{Alternative allocation policy: Fair Queuing}\label{alternative-allocation-policy-fair-queuing}

The short-priority allocation policy (DRR biased toward interactive
classes) optimizes for short-request tail latency at the cost of
potentially starving heavy requests. An alternative design objective is
fairness: giving each class equal service opportunities regardless of
request size. We implement Fair Queuing as a round-robin allocation
between short and heavy classes and compare it against both direct naive
(FIFO) and short-priority scheduling under a heavy-dominated workload
(70\% long/xlong requests).

Table 4 summarizes P90 latency and fairness metrics across the three
policies.

\begin{table}[h]
\centering
\small
\begin{tabular}{lrrr}
\toprule
Policy & Short P90 (ms) & Long P90 (ms) & Global Stdev \\
\midrule
\textbf{Direct (FIFO)} & 10279 & 48481 & 44572 \\
\textbf{Short-Priority} & 7479 (+27\%) & 104638 (-116\%) & 50197 \\
\textbf{Fair Queuing} & 6973 (+32\%) & 56949 (-17\%) & 46879 \\
\bottomrule
\end{tabular}
\caption{Fair Queuing vs Short-Priority allocation (Table 4).}
\end{table}

\textbf{Key observations:}

\begin{enumerate}
\def\labelenumi{\arabic{enumi}.}
\item
  \textbf{Short-request latency}: Fair Queuing achieves +32\%
  improvement over FIFO---slightly better than Short-Priority's
  +27\%---because round-robin still gives short requests timely service
  while avoiding the queue-depth buildup that heavy-first arrivals
  create under FIFO.
\item
  \textbf{Long-request cost}: Short-Priority inflates long-request P90
  by +116\% relative to FIFO as heavy jobs wait while short requests
  drain. Fair Queuing's overhead is only +17\%, a 6× reduction in the
  ``fairness tax'' paid by long requests.
\item
  \textbf{Global variance}: Fair Queuing's latency standard deviation
  (46879 ms) is lower than Short-Priority (50197 ms) and close to FIFO
  (44572 ms), indicating more uniform treatment across request types.
\end{enumerate}

\textbf{Trade-off interpretation:} Short-Priority is appropriate when
interactive latency is the primary objective and heavy-request
starvation is acceptable. Fair Queuing offers a balanced operating
point---short-tail improvements comparable to Short-Priority with
substantially lower impact on heavy requests---suitable for mixed
workloads where both classes have service-level expectations.

This comparison extends the allocation layer design space (§3.1) and
shows that the three-layer decomposition accommodates different fairness
objectives without changing the ordering or overload control components.

\subsection{Overload semantics and shedding-policy evidence}\label{overload-semantics-and-shedding-policy-evidence}

We aggregate overload actions over all Final (OLC) cells in the main
benchmark---four regimes × five seeds = 20 runs (only this strategy's
runs are included). Short requests see zero rejections; medium requests
are admitted without defer/reject; long jobs are mostly deferred under
stress; xlong jobs bear the majority of rejections. Figure 5 shows the
default cost-ladder \texttt{bucket\_policy}.

\begin{figure}
\centering
\pandocbounded{\includegraphics[keepaspectratio,alt={Overload actions summed over Final (OLC) main-benchmark runs (20 runs: four regimes × five seeds): rejections concentrate on xlong; short requests are never rejected.}]{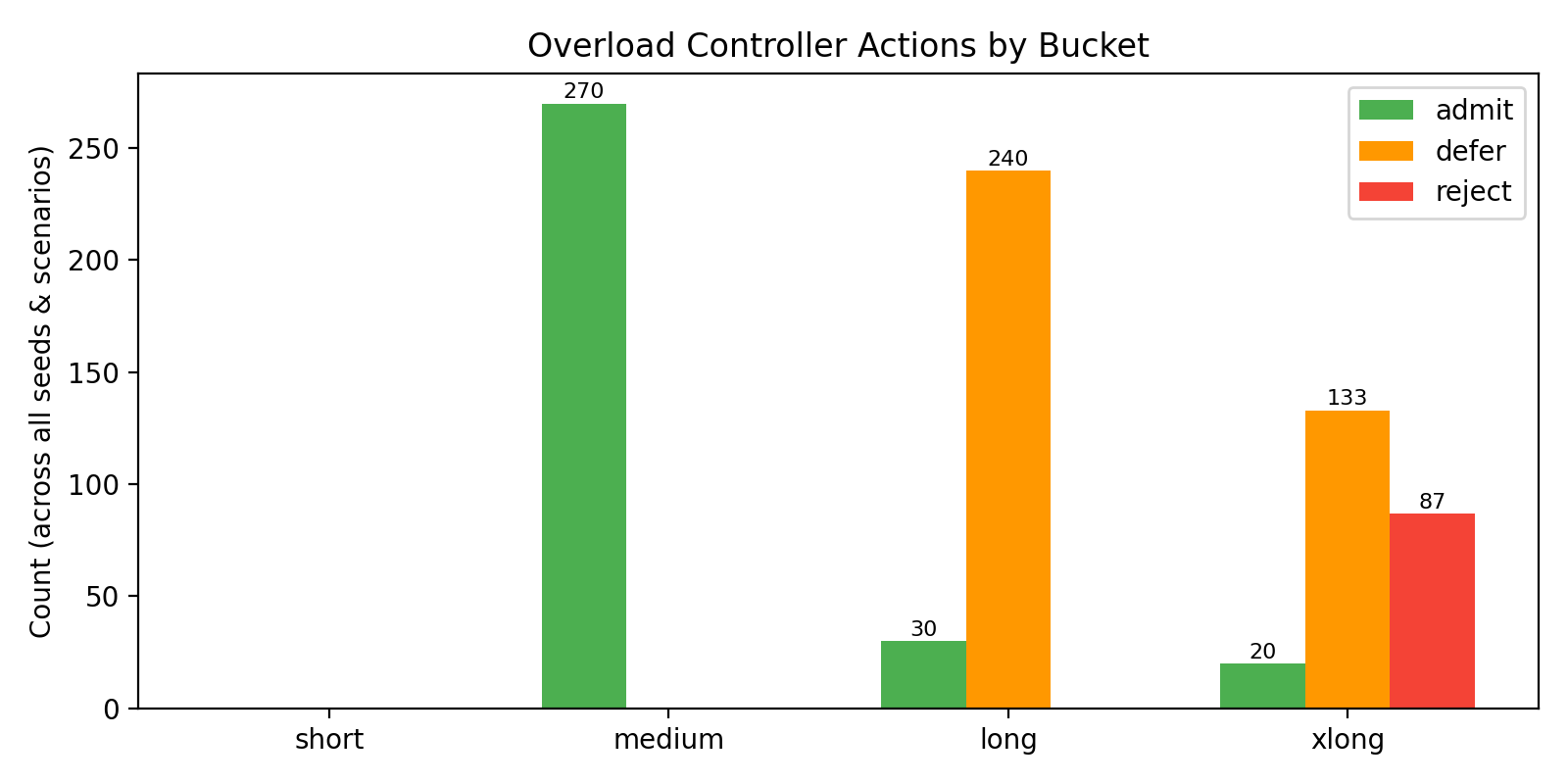}}
\caption{Overload actions summed over Final (OLC) main-benchmark runs
(20 runs: four regimes × five seeds): rejections concentrate on xlong;
short requests are never rejected.}
\end{figure}

Holding Final (OLC) fixed, we vary only
\texttt{overload\_controller.bucket\_policy} under balanced / high and
heavy-dominated / high (five seeds each). Cost ladder (\texttt{ladder})
is the default long/xlong severity map. Uniform mild keeps the same
thresholds but applies one shared mid-tier severity to medium/long/xlong
(class-agnostic among non-short work). Uniform harsh applies the
harshest non-short tier uniformly. Reverse inverts the long/xlong
ordering as a stress contrast only.

Table 5 and Figure 6 summarize mean ± std
(\texttt{paper\_results/tables/overload\_policy\_comparison\_summary.csv}).

\begin{table}[h]
\centering
\footnotesize
\begin{tabular}{llrrrrrrr}
\toprule
Regime & Policy & Short P95 & Global P95 & CR & Satisf. & Goodput & Rejects & Defers \\
\midrule
\textbf{bal./high} & Cost ladder & 342±30 & 7458±1200 & 1.00 & 1.00 & \textbf{4.3±1.5} & 4.8±0.8 & 8.6±1.3 \\
& Uniform mild & 332±12 & 21538±3076 & 1.00 & 1.00 & 1.4±0.2 & 0 & 26.6±1.5 \\
& Uniform harsh & 340±4 & 5299±3576 & 1.00 & 1.00 & 3.9±1.7 & 17.0±4.6 & 9.8±4.4 \\
& Reverse & 353±16 & 21845±3949 & 1.00 & 0.98±0.02 & 1.4±0.1 & 10.4±3.1 & 10.4±2.9 \\
\midrule
\textbf{heavy/high} & Cost ladder & 338±19 & 23303±7459 & 1.00 & 0.93±0.06 & 1.7±0.8 & 14.0±2.8 & 20.0±3.7 \\
& Uniform mild & 328±18 & 44249±2500 & \textbf{0.88±0.01} & 0.81±0.07 & 0.8±0.1 & 0 & 46.0±1.7 \\
& Uniform harsh & 325±12 & 18782±11013 & 1.00 & 0.98±0.03 & \textbf{2.0±1.3} & 29.2±8.3 & 16.4±9.2 \\
& Reverse & 327±31 & 46331±1343 & \textbf{0.88±0.02} & 0.76±0.09 & 0.8±0.1 & 10.6±0.5 & 17.8±0.8 \\
\bottomrule
\end{tabular}
\caption{Overload policy comparison (Table 5).}
\end{table}

In balanced / high, the cost ladder achieves the highest useful goodput
while preserving full completion and deadline satisfaction. Uniform mild
keeps short tails comparable but shifts pressure into mass deferral (no
rejects) and collapses useful goodput---showing that ``gentle''
class-agnostic admission can hide overload in the queue rather than
shape it into legible sacrifice. Reverse degrades satisfaction despite
similar completion, illustrating that which expensive bucket is targeted
matters for SLO-meeting throughput.

In heavy-dominated / high, uniform mild and reverse drop completion and
satisfaction; the cost ladder maintains full completion with sacrifice
concentrated on xlong (Figure 5). Uniform harsh yields higher useful
goodput and a lower global P95 in this cell but drives more rejects
across all non-short classes---an alternative operating point on the
same joint surface, not a refutation of structured shedding. Together,
the comparison supports the design intent: cost-first defer/reject on
the long/xlong ladder aligns with short-tail protection,
completion-conditioned interpretation, and useful goodput under the
stated objective, while making who is sacrificed explicit in policy.

\begin{figure}
\centering
\pandocbounded{\includegraphics[keepaspectratio,alt={Overload bucket\_policy comparison (Final OLC fixed; balanced/high and heavy-dominated/high). Grouped bars: short P95, useful goodput, completion rate for cost ladder, uniform mild/harsh, and reverse (stress contrast). Mean ± std, five seeds.}]{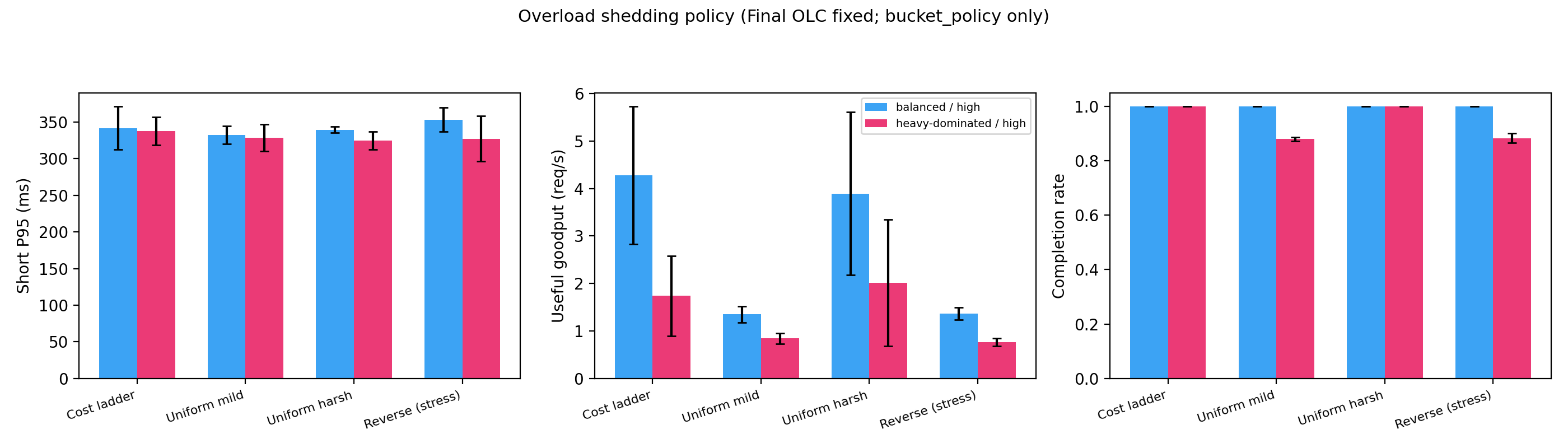}}
\caption{Overload bucket\_policy comparison (Final OLC fixed;
balanced/high and heavy-dominated/high). Grouped bars: short P95, useful
goodput, completion rate for cost ladder, uniform mild/harsh, and
reverse (stress contrast). Mean ± std, five seeds.}
\end{figure}

\subsection{Ablations and layerwise progression}\label{ablations-and-layerwise-progression}

§4.5--4.7 hold the coarse prior fixed and vary either policy family,
allocation fairness, or overload shedding shape. Figure 7 steps through
naive → quota-tiered → adaptive DRR → Final (OLC) on the same two
high-congestion regimes, so layer additions can be read as moves on the
same joint axes.

\begin{figure}
\centering
\pandocbounded{\includegraphics[keepaspectratio,alt={Layerwise progression under high congestion (balanced/high, heavy-dominated/high): short P95, useful goodput, and completion from naive dispatch through quota-tiered isolation, adaptive DRR, and Final (OLC).}]{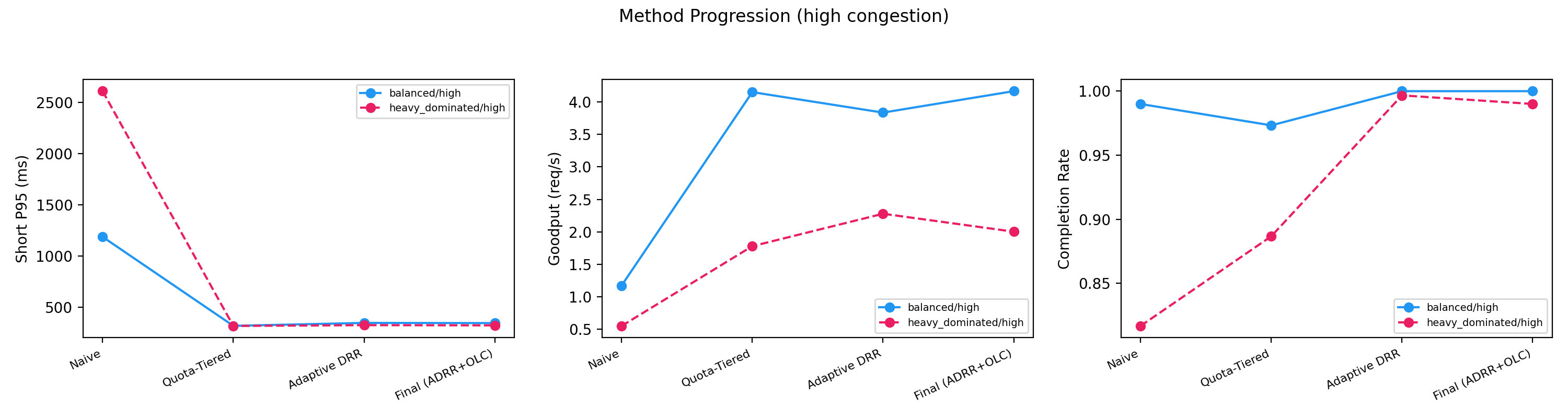}}
\caption{Layerwise progression under high congestion (balanced/high,
heavy-dominated/high): short P95, useful goodput, and completion from
naive dispatch through quota-tiered isolation, adaptive DRR, and Final
(OLC).}
\end{figure}

§4.9 perturbs overload-controller thresholds (defer/reject cutoffs and
backoff) while holding baseline coarse priors fixed. §4.10 perturbs
predictor fidelity---multiplicative error on coarse p50/p90
priors---with the semi-clairvoyant configuration otherwise unchanged.
The two subsections separate controller tuning from numerical prior
quality.

\subsection{Sensitivity analysis}\label{sensitivity-analysis}

After the policy comparisons and layerwise progression in §§4.5--4.8, we
stress overload admission thresholds by perturbing defer/reject cutoffs
and backoff by ±20\% from their baseline values. Completion rate
remained ≥ 0.99 for all variants; deadline satisfaction moved by at most
4.2\%; short-request P95 changed by at most 5.9\% relative to the
baseline configuration. Useful goodput varied with aggressiveness---as
expected when admission becomes stricter or looser---without unstable
collapse. Thresholds are therefore stable under modest perturbation but
not uniquely determined. This subsection isolates controller tuning;
§4.10 isolates predictor fidelity (numerical prior error) with cutoffs
restored to baseline.

\subsection{Sensitivity to predictor quality}\label{sensitivity-to-predictor-quality}

\textbf{Division of labor.} Section §4.4 establishes whether usable
per-request output-length magnitude exists at all (four-level
information ladder under fixed policy). Section §4.9 perturbs overload
thresholds while coarse p50/p90 priors stay at their baseline values.
This subsection assumes semi-clairvoyance is already in place and asks
how much numerical accuracy those priors require when controller cutoffs
are not the moving part. If the three-layer design collapsed as soon as
estimates became noisy, deployability under realistic predictors would
be doubtful.

Semi-clairvoyant client-side scheduling is intentionally framed around
coarse priors rather than oracle knowledge (Gan et al., 2026; §4.4). We
stress predictor quality independently of §4.8 by injecting
deterministic, per-request multiplicative error into the policy-facing
p50/p90 values after the usual coarse prior is formed: each prior is
multiplied by a factor drawn uniformly from {[}1−L, 1+L{]}, with L ∈
\{0, 0.1, 0.2, 0.4, 0.6\} (up to 60\% relative error at the endpoints).
Routing buckets and mock service physics are unchanged---the simulator
still completes each job with the same nominal workload scale---so the
sweep isolates errors in what the client believes about length from
changes in the provider.

We hold the Final (OLC) stack fixed and repeat all four regimes with
five seeds per (L, regime) pair (100 runs total: five values of L × four
regimes × five seeds). Figure 8 plots useful goodput, completion rate,
and short-request P95 versus L.

\begin{figure}
\centering
\pandocbounded{\includegraphics[keepaspectratio,alt={Predictor-quality sweep (Final OLC fixed): multiplicative noise on policy-facing p50/p90 priors with L from 0 to 0.6; mock physics unchanged. Mean ± std over five seeds; one line per regime.}]{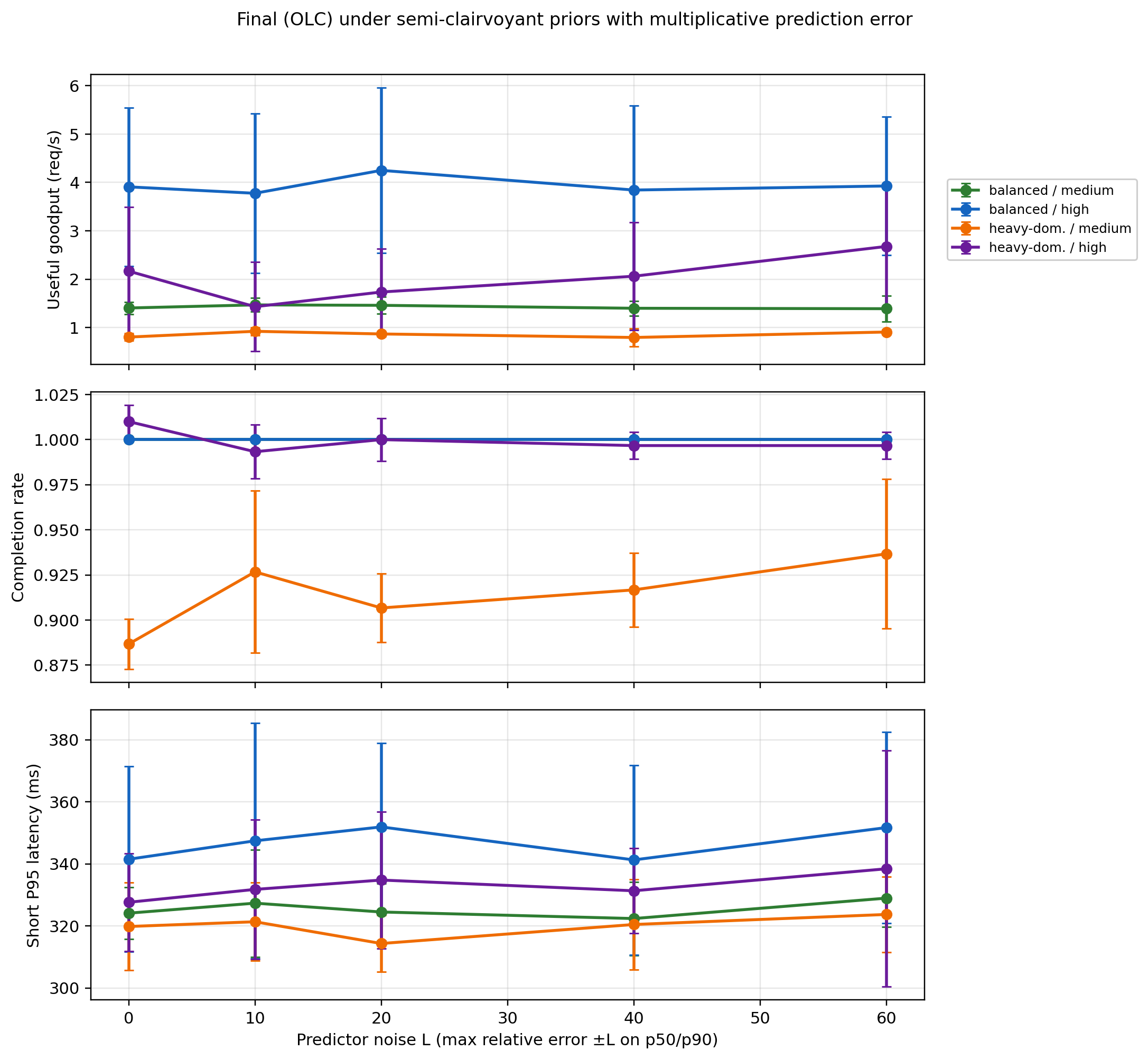}}
\caption{Predictor-quality sweep (Final OLC fixed): multiplicative noise
on policy-facing p50/p90 priors with L from 0 to 0.6; mock physics
unchanged. Mean ± std over five seeds; one line per regime.}
\end{figure}

\textbf{Balanced regimes.} In balanced / high, completion and deadline
satisfaction remain at 1.00 for every L; useful goodput drifts between
roughly 3.8 and 4.3 completed SLO-meeting requests per second without a
sharp cliff---consistent with graceful degradation of the joint
operating point rather than a binary failure when priors are wrong.
Short-request P95 stays in a band near 340--352 ms across noise levels
(within seed variance). Balanced / medium shows the same qualitative
pattern: completion 1.00 across L, short P95 near 324--329 ms, useful
goodput near 1.39--1.47---modest sensitivity to noise relative to the
information ladder in §4.4 (where removing magnitude priors inflates
short P95 by large multiplicative factors in stressed cells).

\textbf{Heavy-dominated regimes.} Heavy-dominated / high keeps
completion near 1.0 (within measurement noise on completion rate) while
short P95 moves from about 328 ms at L = 0 to about 338 ms at L = 0.6
and useful goodput varies between roughly 1.4 and 2.7 with L---again no
abrupt collapse, but stronger coupling to noise than in balanced
workloads, matching the paper's broader theme of regime-dependent
robustness. Heavy-dominated / medium exhibits the largest completion
swing across L (0.89 at L = 0 versus up to 0.94 at L = 0.6) with
deadline satisfaction ranging about 0.80--0.88, reflecting how
mis-estimated token budgets interact with deferral-heavy overload
semantics when the mix is heavy-rich; even here the response is graded,
not a single threshold breakdown.

\textbf{Takeaway.} The experiment supports the layered design under
coarse priors: exact length knowledge is not required for it to remain
actionable. Over a substantial range of multiplicative prediction error,
metrics drift smoothly; severe mis-estimation primarily re-shapes
trade-offs among tails, completion, and useful goodput rather than
invalidating the decomposition outright---complementing §4.4's contrast
between no-information blind, class-only, coarse, and oracle extremes.
Full numerical means and standard deviations are in
\texttt{paper\_results/tables/predictor\_noise\_summary.csv}
(appendix-friendly).

\section{Discussion and
Limitations}\label{discussion-and-limitations}

We study client-side admission and release in front of a black-box API;
in-engine systems (Yu et al., 2022; Kwon et al., 2023; Zhong et al.,
2024; Agrawal et al., 2024; Patel et al., 2024) remain complementary,
not competing, because they control different surfaces.

\textbf{Metrics and evidence.} Useful goodput---finished, SLO-meeting
throughput---follows the same spirit as goodput-oriented generative
serving (Zhong et al., 2024) and SLO-grounded prediction serving
(Crankshaw et al., 2017; Romero et al., 2021) and keeps tail gains from
hiding sacrificed work. §4.4's ladder isolates magnitude from class
structure; §4.9's multiplicative noise sweep holds physics fixed and
shows graded joint-metric drift over L, not a cliff---supporting coarse
rather than oracle priors without claiming calibration realism. The
ShareGPT real-trace validation (§4.1) confirms that policy ordering
holds under a production-derived distribution; full trace replay with
production predictor pipelines remains a natural extension.

\textbf{Overload legibility.} §4.6 contrasts cost-ladder shedding with
uniform and reversed policies: under balanced / high, the ladder
maximizes useful goodput at full completion and satisfaction; under
heavy-dominated / high, it avoids the completion and satisfaction drops
seen for uniform mild and reverse while keeping sacrifice visible at the
client. That separation clarifies policy-induced deferral and rejection
versus failures that appear only as provider-side timeouts.

\textbf{Limits.} The mock is not a vendor model, though its linear
latency scaling (§4.1) is calibrated against production API measurements
(R² = 0.97); five seeds inform trends but leave variance in some cells.
Thresholds were hand-tuned; §4.8's ±20\% sweep shows local stability,
not global optimality. Heavy-dominated settings react more strongly to
predictor noise (§4.9). The structural claim---that allocation,
ordering, and overload are separable client concerns once token priors
exist---does not depend on any single numeric corner.

\textbf{Using the results.} Operators should pick points on the joint
surface (tails, completion, satisfaction, useful goodput). When
completing work matters, the full stack is strong in balanced and
several heavy-dominated cells; when minimizing global tail matters more
than finishing every heavy job, quota-style isolation can win for the
reasons Table 2 already shows. Table 5 illustrates the same idea for
admission shape alone.

\section{Related Work}\label{related-work}

We group prior work by its relationship to our formulation rather than
by exhaustive taxonomy.

\textbf{Enabling premise.} Gan et al.~(2026) study SageSched: LLM
scheduling under demand uncertainty and hybridity, showing that
predicted output structure can drive scheduling decisions. Relationship
to this paper: we treat that finding as the premise that makes
client-side semi-clairvoyance technically meaningful at the
API---without coarse length predictability, the black-box client lacks a
workload unit for allocation and admission.

\textbf{In-engine control plane.} Orca (Yu et al., 2022),
vLLM/PagedAttention (Kwon et al., 2023), DistServe (Zhong et al., 2024),
Sarathi-Serve (Agrawal et al., 2024), and Splitwise (Patel et al., 2024)
optimize batching, memory, and pipelining inside the engine.
Relationship: they define what the client cannot see or reorder; our
question is how to shape arrivals given only API-visible feedback and
coarse priors.

\textbf{Goodput and SLO-oriented serving.} DistServe (Zhong et al.,
2024) argues for goodput as a first-class objective in generative
serving; Clipper and INFaaS (Crankshaw et al., 2017; Romero et al.,
2021) ground latency SLOs and admission in prediction serving.
Relationship: we adopt the same joint view---tails plus completed
SLO-meeting work---but at the black-box client where admission and
class-wise release are the only levers.

\textbf{Fairness and predictability.} GPS-style fair queueing (Parekh
and Gallager, 1993) informs token-weighted allocation; Clockwork
(Gujarati et al., 2020) stresses predictability in deep inference from
the server side. Relationship: we use fair-share structure externally on
a token prior, while Clockwork-style internal predictability remains
unavailable.

\textbf{Commercial API gateways.} Production LLM APIs (OpenAI, Azure,
Anthropic) enforce token-bucket rate limiting---requests consume tokens
from a replenishing bucket; exhaustion triggers HTTP 429 responses.
Azure API Management supports tiered quotas (tokens-per-minute,
requests-per-minute) with burst allowances. Kong AI Gateway and similar
proxies add retry-after backoff and priority queues. Relationship: these
mechanisms operate server-side or at an intermediary; our work
complements them with client-side shaping that can anticipate overload
before the provider rejects, using coarse priors to defer expensive work
proactively rather than reactively.

To our knowledge, prior work does not systematically formulate and
evaluate allocation, ordering, and overload as separable client-side
concerns for semi-clairvoyant black-box LLM APIs; this paper supplies
that decomposition and a controlled map of joint metrics across regimes,
including overload shedding shape.

\section{Conclusion}\label{conclusion}

Given coarse output-length predictability (Gan et al., 2026),
client-side scheduling against a black-box LLM API is a well-posed
semi-clairvoyant problem. We contribute the allocation / ordering /
overload control decomposition---three separable concerns at that
boundary---and a controlled mock study of how they move short-request
tails, completion, deadline satisfaction, and useful goodput together.
Trade-offs are regime-dependent: a strong joint point under balanced /
high congestion, an information ladder that shows coarse magnitude
matters beyond class-only routing, overload-policy evidence for
cost-ladder shedding, and a predictor-noise sweep aligned with
coarse-not-oracle priors.

The allocation layer admits different fairness objectives:
Short-Priority optimizes interactive latency at the cost of
heavy-request starvation (+27\% short P90 improvement, +116\% long P90
overhead), while Fair Queuing achieves a balanced operating point (+32\%
short improvement with only +17\% long overhead) suitable for mixed
workloads with service-level expectations across all classes. This
demonstrates that the three-layer decomposition accommodates policy
alternatives without restructuring the remaining stack.

Trace-driven replay under the ShareGPT distribution (§4.1) confirms
external validity beyond synthetic workloads; shadow deployment with
production predictor pipelines is the natural next step.

\section*{References}\label{references}

\begin{enumerate}
\def\labelenumi{\arabic{enumi}.}
\item
  Zhenghao Gan, Yichen Bao, Yifei Liu, Chen Chen, Quan Chen, and Minyi
  Guo. SageSched: Efficient LLM Scheduling Confronting Demand
  Uncertainty and Hybridity. \emph{arXiv preprint arXiv:2603.07917},
  March 2026.
\item
  G.-I. Yu et al.~Orca: A Distributed Serving System for
  Transformer-Based Generative Models. In \emph{OSDI}, 2022.
\item
  W. Kwon et al.~Efficient Memory Management for Large Language Model
  Serving with PagedAttention. In \emph{SOSP}, 2023.
\item
  Y. Zhong et al.~DistServe: Disaggregating Prefill and Decoding for
  Goodput-Optimized Large Language Model Serving. In \emph{OSDI}, 2024.
\item
  A. Agrawal et al.~Taming Throughput--Latency Tradeoff in LLM Inference
  with Sarathi-Serve. In \emph{OSDI}, 2024.
\item
  P. Patel et al.~Splitwise: Efficient Generative LLM Inference Using
  Phase Splitting. In \emph{NSDI}, 2024.
\item
  A. K. Parekh and R. G. Gallager. A Generalized Processor Sharing
  Approach to Flow Control in Integrated Services Networks: The
  Single-Node Case. \emph{IEEE/ACM Trans. Networking}, 1(3):344--357,
  1993.
\item
  D. Crankshaw et al.~Clipper: A Low-Latency Online Prediction Serving
  System. In \emph{NSDI}, 2017.
\item
  F. Romero et al.~INFaaS: Automated Model-less Inference Serving. In
  \emph{ATC}, 2021.
\item
  A. Gujarati et al.~Serving DNNs like Clockwork: Performance
  Predictability from the Bottom Up. In \emph{OSDI}, 2020.
\end{enumerate}

\end{document}